\documentclass[12pt]{article}
\usepackage{graphicx}

\def\hybrid{\topmargin 0pt      \oddsidemargin 0pt
        \headheight 0pt \headsep 0pt
       \voffset-1cm
        \textwidth 6.25in       
       \textheight 9.5in       
        \marginparwidth 0.0in
        \parskip 5pt plus 1pt   \jot = 1.5ex}
\catcode`\@=11
\def\marginnote#1{}

\newcount\hour
\newcount\minute
\newtoks\amorpm
\hour=\time\divide\hour by60
\minute=\time{\multiply\hour by60 \global\advance\minute by-\hour}
\edef\standardtime{{\ifnum\hour<12 \global\amorpm={am}%
        \else\global\amorpm={pm}\advance\hour by-12 \fi
        \ifnum\hour=0 \hour=12 \fi
        \number\hour:\ifnum\minute<10 0\fi\number\minute\the\amorpm}}
\edef\militarytime{\number\hour:\ifnum\minute<10 0\fi\number\minute}

\def\draftlabel#1{{\@bsphack\if@filesw {\let\thepage\relax
   \xdef\@gtempa{\write\@auxout{\string
      \newlabel{#1}{{\@currentlabel}{\thepage}}}}}\@gtempa
   \if@nobreak \ifvmode\nobreak\fi\fi\fi\@esphack}
        \gdef\@eqnlabel{#1}}
\def\@eqnlabel{}
\def\@vacuum{}
\def\draftmarginnote#1{\marginpar{\raggedright\scriptsize\tt#1}}

\def\draftlabel#1{{\@bsphack\if@filesw {\let\thepage\relax
   \xdef\@gtempa{\write\@auxout{\string
      \newlabel{#1}{{\@currentlabel}{\thepage}}}}}\@gtempa
   \if@nobreak \ifvmode\nobreak\fi\fi\fi\@esphack}
        \gdef\@eqnlabel{#1}}
\def\@eqnlabel{}
\def\@vacuum{}
\def\draftmarginnote#1{\marginpar{\raggedright\scriptsize\tt#1}}

\def\draft{\oddsidemargin -.5truein
        \def\@oddfoot{\sl preliminary draft \hfil
        \rm\thepage\hfil\sl\today\quad\militarytime}
        \let\@evenfoot\@oddfoot \overfullrule 3pt
        \let\label=\draftlabel
        \let\marginnote=\draftmarginnote
   \def\@eqnnum{(\theequation)\rlap{\kern\marginparsep\tt\@eqnlabel}%
\global\let\@eqnlabel\@vacuum}  }

\def\numberbysection{\@addtoreset{equation}{section}
        \def\theequation{\thesection.\arabic{equation}}}

\def\underline#1{\relax\ifmmode\@@underline#1\else
        $\@@underline{\hbox{#1}}$\relax\fi}

\def\titlepage{\@restonecolfalse\if@twocolumn\@restonecoltrue\onecolumn
     \else \newpage \fi \thispagestyle{empty}\c@page\z@
        \def\thefootnote{\fnsymbol{footnote}} }

\def\endtitlepage{\if@restonecol\twocolumn \else  \fi
        \def\thefootnote{\arabic{footnote}}
        \setcounter{footnote}{0}}  
\relax

\numberbysection
\hybrid

\newfont{\Bbb}{msbm10 scaled 1\@ptsize00}
\newfont{\Bbbb}{msbm7 scaled 1\@ptsize00}

\newcommand{\DDD}{\raise-1pt\hbox{$\mbox{\Bbbb D}$}}

\newcommand{\UUU}{\raise-1pt\hbox{$\mbox{\Bbbb U}$}}

\newcommand{\ZZ}{\mbox{\Bbb Z}}
\newcommand{\z}{\raise-1pt\hbox{$\mbox{\Bbbb Z}$}}

\newcommand{\SSS}{\mbox{\Bbb S}}
\newcommand{\sss}{\raise-1pt\hbox{$\mbox{\Bbbb S}$}}

\def\beq{\begin{equation}}
\def\eeq{\end{equation}}
\def\p{\partial}

\newtheorem{lemma-definition}{Lemma-Definition}[section]

\begin{document}

\begin{titlepage}

\title{Tau-function of the B-Toda hierarchy}

\author{V. Prokofev\thanks{
Skolkovo Institute of Science and Technology, 143026, Moscow, Russia,
and
Steklov Mathematical Institute of Russian Academy of Sciences,
Gubkina str. 8, Moscow, 119991, Russian Federation;
e-mail: vadprokofev@gmail.com}
\and
A.~Zabrodin\thanks{
Skolkovo Institute of Science and Technology, 143026, Moscow, Russia and
Steklov Mathematical Institute of Russian Academy of Sciences,
Gubkina str. 8, Moscow, 119991, Russian Federation;
e-mail: zabrodin@itep.ru}}

\date{March 2023}
\maketitle


\begin{abstract}

We continue the study of the B-Toda hierarchy (the Toda lattice with the constraint of type B) which can be regarded as a discretization of the BKP hierarchy. We introduce the tau-function of the B-Toda hierarchy and obtain 
the bilinear equations for it. Examples of soliton tau-functions are 
presented in the explicit form.

\end{abstract}

\end{titlepage}

\vspace{5mm}

%

\tableofcontents

\vspace{5mm}

\section{Introduction}

The 2D Toda lattice hierarchy \cite{UT84}
plays a very important role in the theory of integrable
systems. The commuting flows of the hierarchy 
are parametrized by two infinite sets of complex time
variables ${\bf t}=\{t_1, t_2, t_3, \ldots \}$ 
(``positive times'') and $\bar {\bf t}=\{\bar t_1,
\bar t_2, \bar t_3, \ldots \}$ (``negative times''), 
together with the integer-valued ``zeroth time'' $n\in \ZZ$. 
(For soliton solutions $n$ makes also sense as a continuous variable.)
Hereafter, the bar does not
mean complex conjugation. 
Equations of the hierarchy are differential in the times 
${\bf t}$, $\bar {\bf t}$ and
difference in $n$. 
They can be represented in the Lax form as evolution
equations for two Lax operators $L$, $\bar L$ which are pseudo-difference
operators, i.e., half-infinite sums of integer powers of the shift
operators $e^{\pm \p_n}$ with coefficients depending on $n$ and ${\bf t}$,
$\bar {\bf t}$.
A common solution is provided by the tau-function 
$\tau_n ({\bf t}, \bar {\bf t})$
which satisfies an infinite set of bilinear differential-difference
equations of Hirota type \cite{DJKM83,JM83}. 
In a certain sense, the Toda hierarchy can be regarded
as a discretization of the Kadomtsev-Petviashvili (KP) hierarchy. 

Recently, in the joint work of one of the authors with I.Krichever
\cite{KZ22}, a different version of the Toda hierarchy was introduced
(see also the earlier work 
\cite{GK10}, where a similar hierarchy was suggested as an
integrable discretization of the 
Novikov-Veselov equation). 
It was
called in \cite{KZ22} ``the Toda lattice with constraint of type B'' 
or simply
B-Toda. It is a subhierarchy of the Toda lattice defined by imposing
a constraint to the Lax operators, which in a special (``balanced'')
gauge has the form
\beq\label{int1}
L^{\dag}=(e^{\p_n}-e^{-\p_n})\bar L (e^{\p_n}-e^{-\p_n})^{-1},
\eeq
where $e^{\pm \p_n}$ is the shift operator acting as $e^{\pm \p_n}
f(n)=f(n\pm 1)$ and 
$L^{\dag}$ is the conjugate operator 
(the ${}^{\dag}$-operation is defined as 
$(f(n)\circ e^{\p_n})^{\dag}=e^{-\p_n}\circ f(n)=f(n-1)\circ e^{-\p_n}$). 
As it was proved in \cite{KZ22}, 
this constraint is preserved by the flows $\p_{T_k}=
\p_{t_k}-\p_{\bar t_k}$ and is
destroyed by the flows $\p_{t_k}+\p_{\bar t_k}$, so to define the 
hierarchy one should restrict the times as $t_k+\bar t_k=0$ and 
consider $T_k=\frac{1}{2}(t_k -\bar t_k)$ as independent variables. 

The constraint (\ref{int1}) should be compared with the constraint
\beq\label{int2}
(L^{\rm KP})^{\dag}=-\p_x L^{\rm KP} \p_x^{-1}, \qquad x=t_1
\eeq
which can be imposed to
the pseudo-differential Lax operator $L^{\rm KP}$ 
of the KP hierarchy. The ${}^{\dag}$-operation is defined in this case 
as $(f(x)\circ \p_x)^{\dag}=-\p_x \circ f(x)$.
This constraint (which is preserved by the
``odd time flows'' $\p_{t_{2k+1}}$ and is destroyed by the 
``even time flows'' $\p_{t_{2k}}$) defines the
BKP hierarchy \cite{DJKM81}-\cite{Z21} with independent variables 
${\bf t}_{o}=\{t_1, t_3, t_5, \ldots \}$. The ``even times''
${\bf t}_{e}=\{t_2, t_4, t_6, \ldots \}$ are put equal to 0.
The constraint (\ref{int1}) looks as a difference analogue of
(\ref{int2}), that is why it was called ``the constraint of type B''.

The characterization of the BKP hierarchy in terms of the KP tau-function
$\tau^{\rm KP}({\bf t})$ was obtained in \cite{Z21}. The KP tau-functions
that correspond to solutions of the BKP hierarchy satisfy the condition
\beq\label{bkp}
\begin{array}{c}
\Bigl ( \tau^{\rm KP}(t_1-z^{-1}, -\frac{1}{2}z^{-2}, 
t_3-\frac{1}{3}z^{-3}, -\frac{1}{4}z^{-4}, \ldots )\Bigr )^2
\\ \\
=
\tau^{\rm KP}(t_1, 0, t_3, 0, \ldots )
\tau^{\rm KP}(t_1-2z^{-1}, 0, 
t_3-\frac{2}{3}z^{-3}, 0, \ldots ).
\end{array}
\eeq
Besides, there exists a tau-function $\tau^{\rm BKP}({\bf t}_{o})$ 
of the BKP hierarchy which is a function of the ``odd time variables''
${\bf t}_{o}$ only. It is connected with the KP tau-function by the
relation
\beq\label{bkp2}
\tau^{\rm KP}(t_1, 0, t_3, 0, \ldots )=\Bigl (
\tau^{\rm BKP}(t_1, t_3, \ldots )\Bigr )^2.
\eeq
The function $\tau^{\rm BKP}({\bf t}_{o})$ obeys an infinite set 
of bilinear relations of the Hirota type.

The purpose of this paper is to obtain a similar 
characterization of the B-Toda hierarchy
in terms of the Toda tau-function $\tau_n({\bf t}, \bar {\bf t})$
and introduce the tau-function $\tau^B_n({\bf T})$ of the B-Toda 
hierarchy which is a function of the variables ${\bf T}=\{T_1, T_2, 
T_3, \ldots \}$.
The Toda tau-functions that solve the B-Toda hierarchy should 
satisfy the condition
\beq\label{bkp3}
\begin{array}{l}
\tau_{n}({\bf T}, -{\bf T}-[z^{-1}])\tau_{n-1}({\bf T}, 
-{\bf T}-[z^{-1}])
\\ \\
\phantom{aaaaaaaaaaaaaaa}
=(1-z^{-2})\tau_{n}({\bf T}+[z^{-1}], -{\bf T}-[z^{-1}])
\tau_{n-1}({\bf T}, -{\bf T}),
\end{array}
\eeq
where
$$
\begin{array}{c}
{\bf T}\pm [z^{-1}]=\Bigl \{ T_1\pm z^{-1}, T_2 \pm \frac{1}{2}z^{-2},
T_3 \pm \frac{1}{3}z^{-3}, \ldots \Bigr \}.
\end{array}
$$
One may regard (\ref{bkp3}) as 
the constraint (\ref{int1}) represented as a condition 
for the tau-function. 
It singles out the tau-functions of the Toda lattice that solve the
B-Toda hierarchy.
The tau-function $\tau^B_n({\bf T})$ of the B-Toda hierarchy is
connected with $\tau_n({\bf t}, \bar {\bf t})$ by the relation
\beq\label{bkp4}
\tau_n({\bf T}, -{\bf T})=\tau^B_n({\bf T})\, \tau^B_{n-1}({\bf T}),
\eeq
which is a difference analogue of (\ref{bkp2}). 
The function $\tau^B_n({\bf T})$ obeys an infinite set 
of bilinear relations of the Hirota type.

To avoid a misunderstanding, we should stress 
that our B-Toda hierarchy
is very different from what is called the
Toda lattice of type B in \cite{UT84}. The tau-function of the latter
satisfies the condition
\beq\label{but}
\tau_{n}(t_1, 0, t_3, 0, \ldots ; \bar t_1, 0, \bar t_3, 0, \ldots )=
\tau_{1-n}(t_1, 0, t_3, 0, \ldots ; \bar t_1, 0, \bar t_3, 0, \ldots )
\eeq
rather than (\ref{bkp3}), and the square root of
$\tau_{0}(t_1, 0, t_3, 0, \ldots ; 0, 0, 0, 0, \ldots )$
is a tau-function of the BKP hierarchy. 

The plan of the paper is as follows. In section 2 we remind the reader
the main definitions and facts related to the 2D Toda lattice hierarchy.
In section 3 the B-Toda hierarchy is reviewed and the defining 
constraint is represented as a condition for the Toda lattice 
tau-function. In section 4 the tau-function of the B-Toda hierarchy
is introduced and the integral bilinear equation for it is obtained. 
Section 5 is devoted to examples of soliton solutions. Section 6
contains concluding remarks.

\section{The Toda lattice}

Let us briefly review the 2D Toda lattice hierarchy 
following \cite{UT84}.
The basic objects are two pseudo-difference Lax operators
\beq\label{toda1}
{\cal L}=e^{\p_n}+\sum_{k\geq 0}U_k(n) e^{-k\p_n}, \quad
\bar {\cal L}=c(n)e^{-\p_n}+\sum_{k\geq 0}\bar U_k(n) e^{k \p_n}.
\eeq
Here we use the standard gauge in which
the coefficient in front of $e^{\p_n}$ in ${\cal L}$ is equal to 1.
The coefficient functions $c(n)$, $U_k(n)$, $\bar U_k(n)$
are functions of $n$ and all the times ${\bf t}$, $\bar {\bf t}$.
The evolution equations for the Lax operators (the Lax 
equations) are
\beq\label{toda2}
\begin{array}{l}
\p_{t_m}{\cal L}=[{\cal B}_m, {\cal L}], \quad
\p_{t_m}\bar {\cal L}=[{\cal B}_m, \bar {\cal L}],
\quad {\cal B}_m=({\cal L}^m)_{\geq 0}, 
\\ \\
\p_{\bar t_m}{\cal L}=[\bar {\cal B}_m, {\cal L}], \quad
\p_{\bar t_m}\bar {\cal L}=[\bar {\cal B}_m, \bar {\cal L}],
\quad \bar {\cal B}_m=(\bar {\cal L}^m)_{< 0}.
\end{array}
\eeq
Here and below, given a subset $\SSS \subset \ZZ$, we denote
$\displaystyle{\Bigl (\sum_{k\in \z} U_k e^{k \p_n}\Bigr )_{\sss}=
\sum_{k\in \sss} U_k e^{k \p_n}}$.
For example, 
\beq\label{toda4a}
\begin{array}{l}
{\cal B}_1=e^{\p_n}+U_0(n), \qquad
\bar {\cal B}_1=c(n)e^{-\p_n}.
\end{array}
\eeq
Introducing the function $\varphi =\varphi (n)$ via
\beq\label{toda4}
c(n)=e^{\varphi (n)-\varphi (n-1)},
\eeq
we have from (\ref{toda2}):
\beq\label{toda3}
\p_{t_m}\varphi =({\cal L}^m)_0, \quad
\p_{\bar t_m}\varphi =-(\bar {\cal L}^m)_0.
\eeq

An equivalent formulation of the Toda hierarchy is through the 
Zakharov-Shabat equations
\beq\label{toda8}
\begin{array}{l}
\p_{t_k}{\cal B}_m -\p_{t_m}{\cal B}_k +[{\cal B}_m, {\cal B}_k]=0,
\\ \\
\p_{\bar t_k}{\cal B}_m -\p_{t_m}
\bar {\cal B}_k +[{\cal B}_m, \bar {\cal B}_k]=0,
\\ \\
\p_{\bar t_k}\bar {\cal B}_m -\p_{\bar t_m}
\bar {\cal B}_k +[\bar {\cal B}_m, \bar {\cal B}_k]=0.
\end{array}
\eeq
The first equation of the Toda hierarchy is obtained from the second 
equation
in (\ref{toda8}) at $k=m=1$. 
In terms of the function $\varphi (n)$ 
it has the well-known form
\beq\label{toda5}
\p_{t_1}\p_{\bar t_1}\varphi (n)=e^{\varphi (n)-\varphi (n-1 )}-
e^{\varphi (n+1)-\varphi (n)}.
\eeq

An important role in the theory is played by the dressing operators
${\cal W}$, $\bar {\cal W}$ which are defined by the relations
\beq\label{toda101}
{\cal L}={\cal W}e^{\p_n}{\cal W}^{-1}, \quad
\bar {\cal L}=\bar {\cal W}e^{-\p_n}\bar {\cal W}^{-1}.
\eeq
Clearly, these relations define them only up to multiplication from
the right by a pseudo-difference operator with constant coefficients.
This freedom will be fixed later.
The dressing operators have the general form
\beq\label{toda102}
\begin{array}{l}
\displaystyle{
{\cal W}=1+\sum_{k\geq 1} w_k(n)e^{-k\p_n},}
\\ \\
\displaystyle{
\bar {\cal W}=\bar w_0(n)+\sum_{k\geq 1} \bar w_k(n)e^{k\p_n},
\quad \bar w_0(n)=e^{\varphi (n)}.}
\end{array}
\eeq
We also need the operators $({\cal W}^{\dag})^{-1}$, 
$(\bar {\cal W}^{\dag})^{-1}$. They have the form
\beq\label{toda103}
\begin{array}{l}
\displaystyle{
({\cal W}^{\dag})^{-1}=1+\sum_{k\geq 1} w_k^*(n)e^{k\p_n},}
\\ \\
\displaystyle{
(\bar {\cal W}^{\dag})^{-1}
=\bar w_0^*(n)+\sum_{k\geq 1} \bar w_k^*(n)e^{-k\p_n}, \quad
\bar w_0^*(n)=(\bar w_0(n))^{-1}=e^{-\varphi (n)}.}
\end{array}
\eeq

The wave functions depending on the spectral parameter $z$ 
are defined as
\beq\label{toda104}
\begin{array}{l}
\displaystyle{
\psi_n({\bf t}, \bar {\bf t};z)={\cal W}z^n e^{\xi ({\bf t},z)}=
z^n e^{\xi ({\bf t},z)}\Bigl (1+\sum_{k\geq 1}w_k(n)z^{-k}\Bigr ),
\quad z\to \infty ,}
\\ \\
\displaystyle{
\bar \psi_n({\bf t}, \bar {\bf t};z)=\bar {\cal W}z^n 
e^{\xi (\bar {\bf t},z^{-1})}=
z^n e^{\xi (\bar {\bf t},z^{-1})}
\Bigl (\bar w_0(n)+\sum_{k\geq 1}\bar w_k(n)z^{k}\Bigr ),
\quad z\to 0 ,}
\\ \\
\displaystyle{
\psi_n^*({\bf t}, \bar {\bf t};z)=( {\cal W}^{\dag})^{-1}
z^{-n} 
e^{-\xi ({\bf t},z)}=
z^{-n} e^{-\xi ({\bf t},z)}\Bigl (1+\sum_{k\geq 1}w_k^*(n)z^{-k}\Bigr ),
\quad z\to \infty ,}
\\ \\
\displaystyle{
\bar \psi_n^*({\bf t}, \bar {\bf t};z)=(\bar {\cal W}^{\dag})^{-1}
z^{-n }
e^{-\xi (\bar {\bf t},z^{-1})}=
z^{-n} e^{-\xi (\bar {\bf t},z^{-1})}
\Bigl (\bar w_0^*(n)\! +\! \sum_{k\geq 1}\bar w_k^*(n)z^{k}\Bigr ),
\quad \!\! z\to 0 ,}
\end{array}
\eeq
where
\beq\label{toda105}
\xi ({\bf t},z)=\sum_{k\geq 1}t_k z^k.
\eeq
The wave functions satisfy the linear differential-difference equations
\beq\label{toda106}
\begin{array}{l}
\p_{t_m}\psi_n ={\cal B}_m \psi_n , \quad 
\p_{\bar t_m}\psi_n =\bar {\cal B}_m \psi_n , \quad 
{\cal L}\psi_n =z\psi_n, \quad
\bar {\cal L}\psi_n =z^{-1}\psi_n,
\\ \\
\p_{t_m}\bar \psi_n ={\cal B}_m \bar \psi_n , \quad 
\p_{\bar t_m}\bar \psi_n =\bar {\cal B}_m \bar \psi_n , \quad 
{\cal L}\bar \psi_n =z\bar \psi_n, \quad
\bar {\cal L}\bar \psi_n =z^{-1}\bar \psi_n.
\end{array}
\eeq
The Lax and Zakharov-Shabat equations (\ref{toda2}), (\ref{toda8})
are compatibility conditions for these linear equations. The Toda hierarchy
is encoded in the bilinear integral relation for the wave
functions
\beq\label{toda107}
\oint_{C_{\infty}} \Bigl (\psi_n({\bf t}, \bar {\bf t};z)
\psi^*_{n'}({\bf t}', \bar {\bf t}';z)-
\bar \psi_n({\bf t}, \bar {\bf t};z^{-1})
\bar \psi^*_{n'}({\bf t}', \bar {\bf t}';z^{-1})\Bigr )\frac{dz}{z}=0
\eeq
valid for all $n,n'$, ${\bf t}, {\bf t}'$, $\bar {\bf t}, \bar {\bf t}'$,
where $C_{\infty}$ is a big circle around $\infty$ (this relation just
states that the coefficient in front of
$z^{-1}$ in the Laurent expansion of the expression under the integral 
at infinity is equal to 0).

The common solution of the hierarchy is given by the tau-function
$\tau_n ({\bf t}, \bar {\bf t})$. The wave functions can be expressed
through the tau-function as follows \cite{UT84}:
\beq\label{toda108}
\begin{array}{l}
\displaystyle{
\psi_n({\bf t}, \bar {\bf t};z)=z^{n}e^{\xi ({\bf t}, z)}\, 
\frac{\tau_{n}({\bf t}-[z^{-1}], \bar {\bf t})}{\tau_{n}({\bf t}, 
\bar {\bf t})}},
\\ \\
\displaystyle{
\bar \psi_n({\bf t}, \bar {\bf t};z)=z^{n}e^{\xi (\bar {\bf t}, z^{-1})}\, 
\frac{\tau_{n+1}({\bf t}, \bar {\bf t}-[z])}{\tau_{n}({\bf t}, 
\bar {\bf t})}},
\\ \\
\displaystyle{
\psi_n^*({\bf t}, \bar {\bf t};z)=z^{-n}e^{\xi (-{\bf t}, z)}\, 
\frac{\tau_{n+1}({\bf t}+[z^{-1}], \bar {\bf t})}{\tau_{n+1}({\bf t}, 
\bar {\bf t})}},
\\ \\
\displaystyle{
\bar \psi_n^*({\bf t}, \bar {\bf t};z)=
z^{-n}e^{\xi (-\bar {\bf t}, z^{-1})}\, 
\frac{\tau_{n}({\bf t}, \bar {\bf t}+[z])}{\tau_{n+1}({\bf t}, 
\bar {\bf t})}},
\end{array}
\eeq
where we use the standard notation
\beq\label{toda109}
\begin{array}{l}
{\bf t}\pm [z]=\Bigl \{ t_1\pm z, \, t_2 \pm \frac{1}{2} z^2,\,
t_3 \pm \frac{1}{3} z^3, \, \ldots \Bigr \}.
\end{array}
\eeq
In particular, we have
\beq\label{toda5a}
e^{\varphi (n)}=\frac{\tau_{n+1}}{\tau_n}
\eeq
and the Toda equation (\ref{toda5}) becomes a bilinear equation for the 
tau-function.

The bilinear relation (\ref{toda107}) in terms of the tau-function 
reads
\beq\label{toda110}
\begin{array}{l}
\displaystyle{
\oint_{C_{\infty}} z^{n-n'-1} e^{\xi ({\bf t}-{\bf t}', z)}
\tau_n({\bf t}-[z^{-1}], \bar {\bf t})
\tau_{n'+1}({\bf t}'+[z^{-1}], \bar {\bf t}')dz}
\\ \\
=\displaystyle{
\oint_{C_{0}} z^{n-n'-1} e^{\xi (\bar {\bf t}-\bar {\bf t}', z^{-1})}
\tau_{n+1}({\bf t}, \bar {\bf t}-[z])
\tau_{n'}({\bf t}', \bar {\bf t}'+[z])dz},
\end{array}
\eeq
where $C_0$ is a small circle around $0$. Setting $n'=n-1$,
${\bf t}-{\bf t}'=[\lambda^{-1}]$, $\bar {\bf t}-\bar {\bf t}'=
[\mu^{-1}]$ and taking the residues in (\ref{toda110}), we get the 
equation for the tau-function of the Toda lattice of the 
Hirota-Miwa type:
\beq\label{toda111}
\begin{array}{c}
\tau_n({\bf t}-[\lambda^{-1}], \bar {\bf t})\tau_n({\bf t},
\bar {\bf t}-[\mu^{-1}])-
\tau_n({\bf t}, \bar {\bf t})
\tau_n({\bf t}-[\lambda^{-1}], \bar {\bf t}-[\mu^{-1}])
\\ \\
=\, (\lambda \mu )^{-1}\tau_{n+1}({\bf t},
\bar {\bf t}-[\mu^{-1}])\tau_{n-1}
({\bf t}-[\lambda^{-1}], \bar {\bf t}).
\end{array}
\eeq

There exist other equivalent formulations of the Toda hierarchy
obtained from the one given above by ``gauge 
transformations'' \cite{Takebe1,Takebe2}.
So far we have used the standard gauge in which
the coefficient of the first term of ${\cal L}$ is fixed to be $1$. 
In fact
there is a family of gauge transformations with a function $g=g(n)$
of the form
$$
{\cal L}\to g^{-1}{\cal L}g , \quad \bar {\cal L}\to g^{-1}\bar {\cal L}g,
$$
$$
{\cal B}_n \to g^{-1}{\cal B}_n g -g^{-1}\p_{t_n}g, \quad
\bar {\cal B}_n \to g^{-1}\bar {\cal B}_n g -g^{-1}\p_{\bar t_n}g.
$$
We are interested in another special gauge in which the coefficients
in front of the first terms of the two Lax operators coincide.
Let us denote the Lax operators and the generators of the flows
in this gauge by $L$, $\bar L$, $B_m$, $\bar B_m$ respectively:
\beq\label{toda6}
\begin{array}{c}
L=g^{-1}{\cal L}g, \quad \bar L=g^{-1}\bar {\cal L}g,
\\ \\
B_m =g^{-1}{\cal B}_m g -\p_{t_m}\log g , \quad
\bar B_m =g^{-1}\bar {\cal B}_m g -\p_{\bar t_m}\log g.
\end{array}
\eeq
It is easy to see that the function $g(n)$ is determined from the relation
\beq\label{toda7}
e^{\varphi (n)}=g(n)g(n+1).
\eeq
We call this gauge the balanced gauge.

The gauge transformation changes the normalization of the wave
functions as
\beq\label{toda7a}
\Psi_n = g^{-1}(n)\psi_n , \quad \bar \Psi_n = g^{-1}(n)\bar \psi_n.
\eeq
The new wave functions satisfy the linear equations
\beq\label{toda106a}
\begin{array}{l}
\p_{t_m}\Psi_n =B_m \Psi_n , \quad 
\p_{\bar t_m}\Psi_n =\bar B_m \Psi_n , \quad 
L\Psi_n =z\Psi_n, \quad
\bar L\Psi_n =z^{-1}\Psi_n,
\\ \\
\p_{t_m}\bar \Psi_n =B_m \bar \Psi_n , \quad 
\p_{\bar t_m}\bar \Psi_n =\bar B_m \bar \Psi_n , \quad 
L\bar \Psi_n =z\bar \Psi_n, \quad
\bar L\bar \Psi_n =z^{-1}\bar \Psi_n.
\end{array}
\eeq
Their compatibility conditions are the Lax and Zakharov-Shabat equations
for $L, \bar L$, $B_m, \bar B_m$.

\section{The B-Toda hierarchy}

Let $T$ be the following shift operator:
\beq\label{b1}
T=e^{-\varphi (n)}e^{\p_n}.
\eeq
As is shown in \cite{KZ22}, the Toda lattice admits a constraint
of the form
\beq\label{b2}
(T-T^{\dag})\bar {\cal L}={\cal L}^{\dag}(T-T^{\dag})
\eeq
imposed on the Lax operators. In \cite{KZ22} it was called
the constraint of type B. This constraint is invariant under the
flows $\p_{T_k}=\p_{t_k}-\p_{\bar t_k}$ and is destroyed by the 
flows $\p_{t_k}+\p_{\bar t_k}$. Accordingly, we introduce the 
time variables 
\beq\label{b3}
T_k =\frac{1}{2}(t_k-\bar t_k), \quad y_k =\frac{1}{2}(t_k+\bar t_k)
\eeq
and put $y_k=0$ (i.e., $\bar t_k =-t_k=-T_k$) considering the evolution
in the times ${\bf T}=\{T_1, T_2, T_3, \ldots \}$ only. The corresponding
evolution equations are equations of the B-Toda hierarchy.

The B-Toda hierarchy is most conveniently formulated in the 
balanced gauge. The constraint (\ref{b2}) in this gauge reads
\beq\label{b2a}
(e^{\p_n}-e^{-\p_n})\bar L =L^{\dag}(e^{\p_n}-e^{-\p_n}).
\eeq
The generators of the flows $\p_{T_k}$ are
\beq\label{b2b}
A_k=B_k -\bar B_k,
\eeq
and the linear problems read
\beq\label{b2c}
\p_{T_k}\Psi_n =A_k \Psi_n, \quad \p_{T_k}\bar \Psi_n =A_k \bar \Psi_n.
\eeq
In particular,
\beq\label{b2d}
\p_{T_1}\Psi_n =v_n (\Psi_{n+1}-\Psi_{n-1})
\eeq
with some function $v_n$ 
and the same equation holds for $\bar \Psi_n$.
The compatibility conditions for the linear problems (\ref{b2c})
are the Zakharov-Shabat equations
\beq\label{b2ca}
\p_{T_m}A_k-\p_{T_k}A_m +[A_k, A_m]=0.
\eeq
It is proved in \cite{KZ22} that the difference operators $A_k$ 
are divisible from the right by $e^{\p_n}-e^{-\p_n}$:
\beq\label{b2e}
A_k=C_k (e^{\p_n}-e^{-\p_n}),
\eeq
where $C_k$ is a difference operator (i.e., a linear combination of a 
finite number of shifts $e^{\pm \p_n}$).
The first member of the B-Toda hierarchy is the following 
system of equations
for two unknown functions $v_n$, $f_n$ depending on 
two time variables $T_1$, $T_2$: 
\beq\label{int3}
\left \{\begin{array}{l}
\displaystyle{ \p_{T_1}\log (v_nv_{n+1})=
\frac{f_{n+1}}{v_{n+1}}-\frac{f_n}{v_n}},
\\ \\
\p_{T_2}v_n-\p_{T_1}f_n=2v^2_n(v_{n-1}-v_{n+1}).
\end{array} \right.
\eeq
It is obtained from (\ref{b2ca}) at $k=1$,
$m=2$.

Let us temporarily pass to the standard gauge. 
As is easy to see, in terms of the dressing operators
the constraint (\ref{b2}) reads
\beq\label{b4}
\Bigl [ e^{-\p_n}, \, {\cal W}^{\dag}(T-T^{\dag})\bar {\cal W}\Bigr ]=0,
\eeq
which means that ${\cal W}^{\dag}(T-T^{\dag})\bar {\cal W}$ is a 
(pseudo)difference operator with constant coefficients of the form
$\displaystyle{-e^{-\p_n}+\sum_{k\geq 0}a_k e^{k\p_n}}$. 
The explicit form of this operator depends on the freedom in the 
choice of the dressing operators. We fix this freedom by setting
${\cal W}^{\dag}(T-T^{\dag})\bar {\cal W}=e^{\p_n}-e^{-\p_n}$, i.e.,
\beq\label{b5}
(T-T^{\dag})\bar {\cal W}=({\cal W}^{\dag})^{-1}(e^{\p_n}-e^{-\p_n})
\eeq
or, what is the same,
\beq\label{b5a}
(T-T^{\dag}){\cal W}=(\bar {\cal W}^{\dag})^{-1}(e^{\p_n}-e^{-\p_n}).
\eeq
This choice is motivated by the requirement that the trivial dressing
operators ${\cal W}=\bar {\cal W}=1$ provide a (trivial) solution.

Our next goal is to obtain a characterization of the Toda lattice
tau-functions such that they generate solutions of the B-Toda hierarchy.
Using the definition of the wave functions (\ref{toda104}), we 
read relations (\ref{b5}), (\ref{b5a}) as equations for the wave 
functions:
\beq\label{b6}
(z\! -\! z^{-1})\psi_n^*({\bf T}, -{\bf T};z)=
\frac{\tau_{n-1}({\bf T}, \! -{\bf T})}{\tau_{n}({\bf T}, \! -{\bf T})}\,
\bar \psi_{n-1}({\bf T}, \! -{\bf T};z^{-1})-
\frac{\tau_{n}({\bf T}, \! -{\bf T})}{\tau_{n+1}({\bf T}, \! -{\bf T})}\,
\bar \psi_{n+1}({\bf T}, \! -{\bf T};z^{-1}),
\eeq
\beq\label{b6a}
(z\! -\! z^{-1})\bar \psi_n^*({\bf T}, -{\bf T};z^{-1})=
\frac{\tau_{n}({\bf T}, -{\bf T})}{\tau_{n+1}({\bf T}, -{\bf T})}\,
\psi_{n+1}({\bf T}, -{\bf T};z)-
\frac{\tau_{n-1}({\bf T}, -{\bf T})}{\tau_{n}({\bf T}, -{\bf T})}\,
\psi_{n-1}({\bf T}, -{\bf T};z).
\eeq
In their turn, these equations imply, after plugging
(\ref{toda108}), the following relations for the 
tau-functions:
\beq\label{b7}
\begin{array}{l}
\displaystyle{(1-z^{-2})\tau_n ({\bf T}+[z^{-1}], -{\bf T})
\, =\, \frac{\tau_{n}({\bf T}, -{\bf T})}{\tau_{n-1}({\bf T}, -{\bf T})}\,
\tau_{n-1} ({\bf T}, -{\bf T}-[z^{-1}])}
\\ \\
\phantom{aaaaaaaaaaaaaaaaaaaaaaaaa}\displaystyle{-z^{-2}
\frac{\tau_{n-1}({\bf T}, -{\bf T})}{\tau_{n}({\bf T}, -{\bf T})}\,
\tau_{n+1} ({\bf T}, -{\bf T}-[z^{-1}]),}
\end{array}
\eeq
\beq\label{b7a}
\begin{array}{l}
\displaystyle{(1-z^{-2})\tau_n ({\bf T}, -{\bf T}+[z^{-1}])
\, =\, \frac{\tau_{n}({\bf T}, -{\bf T})}{\tau_{n+1}({\bf T}, -{\bf T})}\,
\tau_{n+1} ({\bf T}-[z^{-1}], -{\bf T})}
\\ \\
\phantom{aaaaaaaaaaaaaaaaaaaaaaaaa}\displaystyle{-z^{-2}
\frac{\tau_{n+1}({\bf T}, -{\bf T})}{\tau_{n}({\bf T}, -{\bf T})}\,
\tau_{n-1} ({\bf T}-[z^{-1}], -{\bf T}).}
\end{array}
\eeq
These relations are specific for the tau-functions of those 
solutions to the Toda lattice that correspond to solutions of the 
B-Toda hierarchy. 
We are going to combine them with the Hirota-Miwa equation (\ref{toda111})
which we write here in the form
\beq\label{b8}
\begin{array}{l}
\tau_{n}({\bf T}+[z^{-1}], -{\bf T})\tau_{n}({\bf T}, -{\bf T}-[z^{-1}])
\, =\, 
\tau_{n}({\bf T}, -{\bf T})\tau_{n}({\bf T}+[z^{-1}], -{\bf T}-[z^{-1}])
\\ \\
\phantom{aaaaaaaaaaaaaaaaaaa}
-\, z^{-2}\tau_{n+1}({\bf T}+[z^{-1}], -{\bf T}-[z^{-1}])
\tau_{n-1}({\bf T}, -{\bf T})
\end{array}
\eeq
(we have set $\lambda =\mu =z$ and shifted ${\bf t}\rightarrow 
{\bf t}+[z^{-1}]$ in (\ref{toda111})).  
Substituting $\tau_{n}({\bf T}+[z^{-1}], -{\bf T})$ from (\ref{b7}) into
(\ref{b8}), we get $\tau_n X_n -z^{-2}\tau_{n-1}X_{n+1}=0$, where
$$
X_n=\frac{\tau_{n}({\bf T}, -{\bf T}-[z^{-1}])
\tau_{n-1}({\bf T}, -{\bf T}-[z^{-1}])}{(1-z^{-2})
\tau_{n-1}({\bf T}, -{\bf T})}-
\tau_{n}({\bf T}+[z^{-1}], -{\bf T}-[z^{-1}]).
$$
Since this has to be satisfied identically, we conclude that $X_n=0$,
i.e,
\beq\label{b9}
\begin{array}{l}
\tau_{n}({\bf T}, -{\bf T}-[z^{-1}])\tau_{n-1}({\bf T}, 
-{\bf T}-[z^{-1}])
\\ \\
\phantom{aaaaaaaaaaaaaaa}
=(1-z^{-2})\tau_{n}({\bf T}+[z^{-1}], -{\bf T}-[z^{-1}])
\tau_{n-1}({\bf T}, -{\bf T}).
\end{array}
\eeq
The same relation follows from (\ref{b7a}). One may regard it as 
the constraint (\ref{b2}) written in terms of the tau-function. 
It singles out the tau-functions of the Toda lattice that solve the
B-Toda hierarchy. Expanding (\ref{b9}) in powers of $z^{-1}$, we obtain,
in the first non-vanishing order:
\beq\label{b10}
\Bigl (\p_{t_1}\log \tau_n +\p_{\bar t_1}\log \tau_{n-1}\Bigr )
\Bigr |_{\, \bar {\bf t}=-{\bf t}}=0.
\eeq
We conjecture that the more general condition (\ref{b9}) already
follows from (\ref{b10}), as it is true for the CKP hierarchy \cite{KZ21}. 

\section{Tau-function of the B-Toda hierarchy}

So far we dealt with the Toda lattice tau-function
$\tau_n({\bf t}, \bar {\bf t})$, which depends on
both sets of times ${\bf t}$ and $\bar {\bf t}$. It turns out
that there exists another tau-function, $\tau_n^B({\bf T})$, which
depends on the times $T_k=\frac{1}{2}(t_k-\bar t_k)$ 
only and which may be called the 
tau-function of the B-Toda hierarchy. The two tau-functions are 
connected by the relation
\beq\label{tau1}
\tau_n({\bf T}, -{\bf T})=\tau_n^B({\bf T})\, \tau_{n-1}^B({\bf T}),
\eeq
where it is implied that the tau-function in the left hand side
satisfied the condition (\ref{b9}).
This condition then states that
\beq\label{tau2}
\tau_n({\bf T}-[z^{-1}], -{\bf T})=(1\! -\! z^{-2})^{1/2}\tau_n^B({\bf T})\, 
\tau_{n-1}^B({\bf T}-[z^{-1}])
\eeq
and
\beq\label{tau2a}
\tau_n({\bf T}, -{\bf T}-[z^{-1}])=(1\! -\! z^{-2})^{1/2}
\tau_n^B({\bf T}+[z^{-1}])\, 
\tau_{n-1}^B({\bf T}).
\eeq
At $z=\infty$ we get (\ref{tau1}). Equations (\ref{b7}), (\ref{b7a})
become:
\beq\label{tau3}
(1\! -\! z^{-2})^{1/2}
\tau_n({\bf T}+[z^{-1}], -{\bf T})=\tau_{n-1}^B({\bf T}+[z^{-1}])
\tau_{n}^B({\bf T})-z^{-2}\tau_{n+1}^B({\bf T}+[z^{-1}])
\tau_{n-2}^B({\bf T}),
\eeq
\beq\label{tau3a}
(1\! -\! z^{-2})^{1/2}
\tau_n({\bf T}, -{\bf T}+[z^{-1}])=\tau_{n-1}^B({\bf T})
\tau_{n}^B({\bf T}-[z^{-1}])-z^{-2}\tau_{n+1}^B({\bf T})
\tau_{n-2}^B({\bf T}-[z^{-1}]).
\eeq
Substituting equations (\ref{tau2})--(\ref{tau3a}) into the bilinear
relation (\ref{toda110}), we get:
\beq\label{tau4}
\begin{array}{l}
\displaystyle{
\tau_{n'+1}^B({\bf T}')\oint_{C_{\infty}}
\Bigl [ z^{n-n'-1}e^{\xi ({\bf T}-{\bf T}',z)}
\tau_{n-1}^B({\bf T}-[z^{-1}])\tau_{n'}^B({\bf T}'+[z^{-1}])}
\\ \\
\displaystyle{\phantom{aaaaaaaaaaaaaaaaaa}
+z^{n'-n-3}e^{-\xi ({\bf T}-{\bf T}',z)}
\tau_{n+1}^B({\bf T}+[z^{-1}])\tau_{n'-2}^B({\bf T}'-[z^{-1}])\Bigr ]dz}
\\ \\
\displaystyle{
=\, \tau_{n'-1}^B({\bf T}')\oint_{C_{\infty}}
\Bigl [ z^{n-n'-3}e^{\xi ({\bf T}-{\bf T}',z)}
\tau_{n-1}^B({\bf T}-[z^{-1}])\tau_{n'+2}^B({\bf T}'+[z^{-1}])}
\\ \\
\displaystyle{\phantom{aaaaaaaaaaaaaaaaaa}
+z^{n'-n-1}e^{-\xi ({\bf T}-{\bf T}',z)}
\tau_{n+1}^B({\bf T}+[z^{-1}])\tau_{n'}^B({\bf T}'-[z^{-1}])\Bigr ]dz}.
\end{array}
\eeq
On the first glance, this looks like a set of 
equations which are cubic in $\tau_n^B$
rather than bilinear. However, below we show that these equations 
are actually equivalent to bilinear equations. 

Indeed, (\ref{tau4}) has the form
\beq\label{tau4a}
G_{n, n'}({\bf T}, {\bf T}')-G_{n, n'+2}({\bf T}, {\bf T}')=0,
\eeq
where 
\beq\label{tau5}
\begin{array}{l}
\displaystyle{
G_{n, n'}({\bf T}, {\bf T}')=\frac{1}{2\pi i}
\oint_{C_{\infty}} \left [ z^{n-n'-1}e^{\xi ({\bf T}-{\bf T}',z)}
\frac{\tau^B_{n-1}({\bf T}-[z^{-1}])}{\tau^B_{n}({\bf T})}\,
\frac{\tau^B_{n'}({\bf T}'+[z^{-1}])}{\tau^B_{n'-1}({\bf T}')}\right. }
\\ \\
\displaystyle{\phantom{aaaaaaaaaaaaaaaa}
\left. +\, z^{n'-n-3}e^{-\xi ({\bf T}-{\bf T}',z)}
\frac{\tau^B_{n+1}({\bf T}+[z^{-1}])}{\tau^B_{n}({\bf T})}\,
\frac{\tau^B_{n'-2}({\bf T}'-[z^{-1}])}{\tau^B_{n'-1}({\bf T}')}\right ]
dz.}
\end{array}
\eeq
Equation (\ref{toda7}) that defines the balanced gauge implies that
\beq\label{tau6}
g(n)=\frac{\tau^B_n({\bf T})}{\tau^B_{n-1}({\bf T})}.
\eeq
Therefore, the wave functions in the balanced gauge are
\beq\label{tau7}
\Psi_n({\bf T}; z)=\frac{\tau^B_{n-1}({\bf T})}{\tau^B_{n}({\bf T})}\,
\psi_n({\bf T}, -{\bf T};z) =(1-z^{-2})^{1/2}
z^n e^{\xi ({\bf T},z)}\, 
\frac{\tau^B_{n-1}({\bf T}-[z^{-1}])}{\tau^B_{n}({\bf T})},
\eeq
\beq\label{tau7a}
\bar \Psi_n({\bf T}; z)=\frac{\tau^B_{n-1}({\bf T})}{\tau^B_{n}({\bf T})}\,
\bar \psi_n({\bf T}, -{\bf T};z^{-1}) =(1-z^{-2})^{1/2}
z^{-n} e^{-\xi ({\bf T},z)}\, 
\frac{\tau^B_{n+1}({\bf T}+[z^{-1}])}{\tau^B_{n}({\bf T})}.
\eeq
Using (\ref{tau7}), (\ref{tau7a}), we can write (\ref{tau5}) in the form
\beq\label{tau8}
G_{n, n'}({\bf T}, {\bf T}')=\frac{1}{2\pi i}
\oint_{C_{\infty}} \Bigl [ \Psi_n({\bf T};z)\bar \Psi_{n'-1}({\bf T}';z)
+\bar \Psi_n({\bf T};z)\Psi_{n'-1}({\bf T}';z)\Bigr ] \frac{dz}{z^2-1}.
\eeq
Equation (\ref{tau4a}) states that $G_{n, n'}({\bf T}, {\bf T}')$ is
periodic in $n'$ with period 2. Therefore, it is enough to consider the
cases $n'=n$ and $n'=n+1$. Let us first put
${\bf T}'={\bf T}$. It is easily seen from (\ref{tau5}) that
$G_{n, n}({\bf T}, {\bf T})=1$, $G_{n, n+1}({\bf T}, {\bf T})=0$,
i.e., $G_{n, n'}({\bf T}, {\bf T})=\frac{1}{2}(1\! +\! (-1)^{n-n'})$.
The following argument shows that $G_{n, n'}({\bf T}, {\bf T}')$
is in fact independent of ${\bf T}'$, i.e.,
$G_{n, n'}({\bf T}, {\bf T}')=\frac{1}{2}(1\! +\! (-1)^{n-n'})$ for all
${\bf T}$, ${\bf T}'$. Let us represent $\bar \Psi_{n'-1}({\bf T}';z)$ in (\ref{tau8}) as a multi-variable Taylor series in powers of
${\bf T}'-{\bf T}$. Each term of the series contains multiple derivatives
of $\bar \Psi_{n'-1}({\bf T};z)$ with respect to the times $T_k$. 
Equations (\ref{b2c}) imply that the action of such a differential operator
to $\bar \Psi_{n'-1}({\bf T};z)$ is equivalent to the action of a
difference operator in $n'$. Moreover, as it follows from (\ref{b2e}),
this difference operator is of the form $D (e^{\p_{n'}}-e^{-\p_{n'}})$
with some operator $D$. The same is true
for the Taylor series for
$\Psi_{n'-1}({\bf T}';z)$. So each term in the expansion of 
$G_{n, n'}({\bf T}, {\bf T}')$ (except the very first one) is of the form
$D (e^{\p_{n'}}-e^{-\p_{n'}})G_{n, n'}({\bf T}, {\bf T}')$  
which i equal to $0$ because
of (\ref{tau4a}). Therefore, we have obtained the bilinear relation
for the wave functions
\beq\label{tau9}
\frac{1}{\pi i}
\oint_{C_{\infty}} \Bigl [ \Psi_n({\bf T};z)\bar \Psi_{n'}({\bf T}';z)
+\bar \Psi_n({\bf T};z)\Psi_{n'}({\bf T}';z)\Bigr ] \frac{dz}{z^2-1}=
1\! -\! (-1)^{n-n'}.
\eeq
In terms of the tau-function it reads
\beq\label{tau10}
\begin{array}{l}
\displaystyle{
\frac{1}{\pi i}
\oint_{C_{\infty}} \Bigl [z^{n-n'-2}e^{\xi ({\bf T}-{\bf T}',z)}
\tau^B_{n-1}({\bf T}-[z^{-1}])\tau^B_{n'+1}({\bf T}'+[z^{-1}])}
\\ \\
\displaystyle{\phantom{aaaaaaaa}
+\, z^{n'-n-2}e^{-\xi ({\bf T}-{\bf T}',z)}
\tau^B_{n+1}({\bf T}+[z^{-1}])\tau^B_{n'-1}({\bf T}'-[z^{-1}])\Bigr ]dz}
\\ \\
\displaystyle{\phantom{aaaaaaaaaaaaaaaaaaaaaaaaaaaaaaaaaa}
=(1\! -\! (-1)^{n-n'})
\tau^B_{n}({\bf T})\tau^B_{n'}({\bf T}').}
\end{array}
\eeq
It is valid for all $n, n'$ and ${\bf T}$, ${\bf T}'$. It is 
not difficult to
see that the simplest ($n$-independent) solution is
$$
\tau^B_n({\bf T})=\exp \Bigl (\frac{1}{2}\sum_{k\geq 1}kT_k^2\Bigr ).
$$
Other examples of solutions are given in the next section.

Setting $n'=n-1$, ${\bf T}-{\bf T}'=[\lambda^{-1}]+[\mu^{-1}]$, 
one can calculate the integral in (\ref{tau10}) by the residue 
calculus. This yields the following 4-term bilinear equation
of the Hirota-Miwa type:
\beq\label{tau11}
\begin{array}{c}
\lambda \tau^B_{n-1}({\bf T}-[\mu^{-1}])
\tau^B_{n}({\bf T}-[\lambda^{-1}])-\mu
\tau^B_{n-1}({\bf T}-[\lambda^{-1}])
\tau^B_{n}({\bf T}-[\mu^{-1}])
\\ \\
-(\lambda \! -\! \mu )
\tau^B_{n-1}({\bf T}-[\lambda^{-1}]-[\mu^{-1}])
\tau^B_{n}({\bf T})  =
(\lambda^{-1}\! -\! \mu^{-1})
\tau^B_{n-2}({\bf T}\! -\! [\lambda^{-1}]\! -\! [\mu^{-1}])
\tau^B_{n+1}({\bf T}).
\end{array}
\eeq
Let us show that this equation is equivalent to the fully discrete
BKP equation first appeared in \cite{Miwa82}. Let $p_1, p_2, p_3$
be three discrete integer-valued 
variables. Introduce the function
$$\tau (p_1, p_2, p_3)=(1-\lambda^{-1}\mu^{-1})^{p_1p_2}
\tau^B_{p_1+p_2+p_3+n-2}({\bf T}+p_1[\mu^{-1}]
+p_2 [\lambda^{-1}])$$
and set
$$
\lambda =\frac{c+b}{c-b}, \quad \mu =\frac{c+a}{c-a}.
$$
Then the equation (\ref{tau11}) acquires the form
\beq\label{tau12}
\begin{array}{l}
(c+a)(b+a)(b-c)\tau (p_1+1, p_2, p_3)\tau (p_1, p_2+1, p_3+1)
\\ \\
\phantom{aaaaa}
+(c+b)(b+a)(c-a)\tau (p_1, p_2+1, p_3)\tau (p_1+1, p_2, p_3+1)
\\ \\
\phantom{aaaaaaaa}
+(c+b)(c+a)(a-b)\tau (p_1, p_2, p_3+1)\tau (p_1+1, p_2+1, p_3)
\\ \\
\phantom{aaaaaaaaaaa}
+(c-a)(c-b)(b-a)\tau (p_1, p_2, p_3)\tau (p_1+1, p_2+1, p_3+1)\, =\, 0.
\end{array}
\eeq
which is the discrete BKP equation from \cite{Miwa82}. The
simplest solution is $\tau (p_1, p_2, p_3)=1$. We see that
in the fully discrete setting the BKP and B-Toda equations are basically
the same (they differ only by a linear change of variables). This is
parallel to what happens with the KP and Toda hierarchies: on the 
fully discrete level they become basically the same. 

It is easy to see that equation (\ref{tau11}) contains all linear
equations for the $\Psi$-function of the form (\ref{b2c}). Indeed, 
in terms of the $\Psi$-function (\ref{tau7}) it reads
\beq\label{tau11a}
(1-e^{-\p_n})\Psi_{n+1}({\bf T}, \lambda )e^{-D(\lambda )}
\Psi_{n}({\bf T}, \mu )=\Psi_{n+1}({\bf T}, \lambda )
\Psi_{n}({\bf T}, \mu )-\Psi_{n}({\bf T}, \lambda )
\Psi_{n+1}({\bf T}, \mu ),
\eeq
where $D(\lambda )$ is the differential operator
$$
D(\lambda )=\sum_{k\geq 1}\frac{\lambda^{-k}}{k}\, \p_{T_k}.
$$
Inverting the difference operator $1-e^{-\p_n}$ as
$\displaystyle{(1-e^{-\p_n})^{-1}=\sum_{k\geq 0}e^{-k\p_n}}$, we rewrite
(\ref{tau11a}) in the form
\beq\label{tau11b}
\Bigl (e^{-D(\lambda )}-1\Bigr )\Psi_{n}({\bf T}, \mu )=
\frac{1}{\Psi_{n+1}({\bf T}, \lambda )}\sum_{k\geq 0}
e^{-k\p_n}\Psi_{n}({\bf T}, \lambda )\Bigl (e^{-\p_n}-e^{\p_n}\Bigr )
\Psi_{n}({\bf T}, \mu ),
\eeq
or
\beq\label{tau11c}
\Bigl (e^{-D(\lambda )}-1\Bigr )\Psi_{n}({\bf T}, \mu )=
\sum_{k\geq 0}\frac{\Psi_{n-k}({\bf T}, 
\lambda )}{\Psi_{n+1}({\bf T}, \lambda )}\Bigl (
\Psi_{n-k-1}({\bf T}, \mu )-\Psi_{n-k+1}({\bf T}, \mu )\Bigr ).
\eeq
This is a generating functional expression for the auxiliary
linear problems.
Expanding both sides of this equation in inverse powers of $\lambda$,
we see that the action of differential operators in $T_k$ is 
equivalent to action of difference operators such that they are
divisible by $e^{\p_n}-e^{-\p_n}$ from the right. This is in agreement
with (\ref{b2c}), (\ref{b2e}).

Next we show how equations (\ref{int3}) can be resolved 
in terms of the tau-function $\tau^B_n({\bf T})$. We set \cite{KZ22}:
\beq\label{tau13}
v_n=\frac{\tau^B_{n+1}\tau^B_{n-1}}{(\tau^B_n)^2}, \quad
f_n=\frac{\tau^B_{n+1}\tau^B_{n-1}}{(\tau^B_n)^2}\, \,
\p_{T_1}\! \log \frac{\tau^B_{n+1}}{\tau^B_{n-1}},
\eeq
then the first equation in (\ref{int3}) is satisfied identically
while the second one reads
$$
\frac{\p_{T_2}\tau^B_{n+1}}{\tau^B_{n+1}}+
\frac{\p_{T_2}\tau^B_{n-1}}{\tau^B_{n-1}}-2
\frac{\p_{T_2}\tau^B_{n}}{\tau^B_{n}}
-\frac{\p_{T_1}^2\tau^B_{n+1}}{\tau^B_{n+1}}
+\frac{\p_{T_1}^2\tau^B_{n-1}}{\tau^B_{n-1}}
$$
$$
+2\frac{\p_{T_1}\tau^B_{n+1}\p_{T_1}\tau^B_{n}}{\tau^B_{n+1}\tau^B_{n}}-
2\frac{\p_{T_1}\tau^B_{n-1}\p_{T_1}\tau^B_{n}}{\tau^B_{n-1}\tau^B_{n}}
+2\frac{\tau^B_{n+2}\tau^B_{n-1}}{\tau^B_{n+1}\tau^B_{n}}
-2\frac{\tau^B_{n-2}\tau^B_{n+1}}{\tau^B_{n-1}\tau^B_{n}}=0.
$$
This is a qubic equation in $\tau^B_n$. However, it is easy to see 
that it reduces to the 
bilinear equation
\beq\label{tau14}
\frac{\p_{T_2}\tau^B_{n+1}}{\tau^B_{n+1}}-
\frac{\p_{T_2}\tau^B_{n}}{\tau^B_{n}}
-\frac{\p_{T_1}^2\tau^B_{n+1}}{\tau^B_{n+1}}
-\frac{\p_{T_1}^2\tau^B_{n}}{\tau^B_{n}}
+2\frac{\p_{T_1}\tau^B_{n+1}\p_{T_1}\tau^B_{n}}{\tau^B_{n+1}\tau^B_{n}}
+2\frac{\tau^B_{n+2}\tau^B_{n-1}}{\tau^B_{n+1}\tau^B_{n}}=2.
\eeq

Finally, we will prove that the bilinear relation for the wave
functions (\ref{tau9}) implies existence of the 
tau-function $\tau^B_n({\bf T})$ such that the wave functions are
expressed through it via formulas (\ref{tau7}), (\ref{tau7a}). 
To this end, we write the wave functions in the form
$$
\Psi_n({\bf T};z)=(1\! -\! z^{-2})^{1/2}
z^n e^{\xi ({\bf T}, z)}w_n({\bf T}, z),
$$
$$
\bar \Psi_n({\bf T};z)=(1\! -\! z^{-2})^{1/2}
z^{-n} e^{-\xi ({\bf T}, z)}\bar w_n({\bf T}, z),
$$
where $w_n({\bf T}, z)$, $\bar w_n({\bf T}, z)$ are regular functions 
of $z$ in a neighborhood of $\infty$. Let us put ${\bf T}-{\bf T}'=
[a^{-1}]$ in (\ref{tau9}),
so that
$\displaystyle{e^{\xi ({\bf T}-{\bf T}', z)} =a/(a-z)}$
and consider the cases $n'=n$ and $n'=n-1$. The residue calculus yields
\beq\label{tau15}
\left \{
\begin{array}{l}
w_n({\bf T}, a)\bar w_n({\bf T}-[a^{-1}], a)=
\bar w_n({\bf T}, \infty )w_n({\bf T}-[a^{-1}], \infty ),
\\ \\
w_n({\bf T}, a)\bar w_{n-1}({\bf T}-[a^{-1}], a)=1.
\end{array}
\right.
\eeq
From these equations we conclude that
\beq\label{tau16}
\frac{w_n({\bf T}, a)}{w_{n+1}({\bf T}, a)}=
\frac{w_n({\bf T}-[a^{-1}], \infty )}{w_{n+1}({\bf T}, \infty )}.
\eeq
Consider the function 
$$
\tilde w_n({\bf T}, a)=\frac{w_n({\bf T}, a)}{w_{n}({\bf T}, \infty )},
\qquad \tilde w_n({\bf T}, \infty )=1.
$$
In terms of this function (\ref{tau16}) means that
$$
\log \tilde w_n({\bf T}, a)-\log \tilde w_{n+1}({\bf T}, a)=
\log w_n({\bf T}-[a^{-1}], \infty )-\log w_n({\bf T}, \infty ),
$$
whence 
$$
\log \tilde w_n({\bf T}, a)=F_n({\bf T}-[a^{-1}])-F_n({\bf T})
$$
for some function $F_n$. This means that $w_n({\bf T}, a)$ can be 
represented in the form
$$
w_n({\bf T}, a)=\frac{\rho_n({\bf T}-[a^{-1}])}{\rho_n({\bf T})}\,
w_n({\bf T}, \infty )
$$
with some $\rho_n$. 
Putting 
$$
\tau_n^B({\bf T})=\frac{\rho_n({\bf T})}{w_n({\bf T}, \infty )},
\quad
\tau_{n-1}^B({\bf T})=\rho_n({\bf T}),
$$
we arrive at (\ref{tau7}). Equation (\ref{tau7a}) is obtained from it
using the second equation in (\ref{tau15}). 

\section{Soliton solutions}

$N$-soliton solutions to the B-Toda hierarchy are specializations 
of $2N$-soliton solutions to the 2D Toda lattice.
The general Toda lattice tau-function 
for $2N$-soliton solution has $6N$ arbitrary parameters
$\alpha_i$, $p_i$, $q_i$ ($i=1, \ldots , 2N$) and is given by
\beq\label{ms1}
\begin{array}{c}
\displaystyle{
\tau_n \left [ \begin{array}{c}\alpha_1 \\ p_1, q_1\end{array};
\begin{array}{c}\alpha_2 \\ p_2, q_2\end{array};
\begin{array}{c}\alpha_3 \\ p_3, q_3\end{array};
\begin{array}{c}\alpha_4 \\ p_4, q_4\end{array}; \, \cdots \, ;
\begin{array}{c}\alpha_{2N-1} \\ p_{2N-1}, q_{2N-1}\end{array};
\begin{array}{c}\alpha_{2N} \\ p_{2N}, q_{2N}\end{array}\right ]
({\bf t}, \bar {\bf t})}
\\ \\
\displaystyle{=\exp \Bigl (-\sum_{k\geq 1} kt_k \bar t_k \Bigr )
\det_{1\leq i,j\leq 2N}\left ( \delta_{ij}+
\alpha_i \, \frac{p_i-q_i}{p_i-q_j}\, \Bigl (\frac{p_i}{q_i}\Bigr )^n
e^{\zeta (p_i, q_i; {\bf t}, \bar {\bf t})}
\right ),}
\end{array}
\eeq 
where
$$
\zeta (p_i, q_i; {\bf t}, \bar {\bf t})=\xi ({\bf t}, p_i)-
\xi ({\bf t}, q_i)+\xi (\bar {\bf t}, p_i^{-1})-
\xi (\bar {\bf t}, q_i^{-1}).
$$
Introducing the tau-function
\beq\label{s1}
\tilde \tau_n({\bf t}, \bar {\bf t})=
\exp \Bigl (\sum_{k\geq 1} kt_k \bar t_k \Bigr )
\tau_n({\bf t}, \bar {\bf t}),
\eeq
we have, expanding the determinant in (\ref{ms1}):
\beq\label{s2}
\begin{array}{c}
\displaystyle{
\tilde \tau_n({\bf t}, \bar {\bf t})=\sum_{m=0}^{2N}\,
\sum_{i_1<i_2< \ldots < i_m}^{2N} \Bigl (\prod_{\gamma =1}^m 
\alpha_{i_{\gamma}}e^{\eta_{i_{\gamma}}}\Bigr )
\prod_{\mu <\nu}^{m}c_{i_{\mu}i_{\nu}}}
\\ \\
\displaystyle{
=1+\sum_{i=1}^{2N}\alpha_i e^{\eta_i}+\sum_{i<j}^{2N}\alpha_i \alpha_j
c_{ij}e^{\eta_i +\eta_j}+
\sum_{i<j<k}^{2N}\alpha_i \alpha_j \alpha_k 
c_{ij}c_{ik}c_{jk}e^{\eta_i +\eta_j+\eta_k}+\ldots }\,\, ,
\end{array}
\eeq
where
$\displaystyle{
e^{\eta_i}=\Bigl (\frac{p_i}{q_i}\Bigr )^n
e^{\zeta (p_i, q_i; {\bf t}, \bar {\bf t})}}
$ and
\beq\label{s3}
c_{ij}=\frac{(p_i-p_j)(q_i-q_j)}{(p_i-q_j)(q_i-p_j)}.
\eeq

In order to obtain soliton solutions of the B-Toda hierarchy,
one should specialize (\ref{ms1}) as
\beq\label{s4}
\begin{array}{l}
\displaystyle{
\tau_n \left [ 
\begin{array}{c}\beta_1(q_1^{-1}\! -\! q_1 )
\\ p_1, \, q_1^{-1}\end{array};
\begin{array}{c}\beta_1(p_1\! -\! p_1^{-1} ) \\ q_1, \, p_1^{-1}
\end{array};
\begin{array}{c}\beta_2(q_2^{-1}\! -\! q_2 )
\\ p_2, \, q_2^{-1}\end{array};
\begin{array}{c}\beta_2(p_2\! -\! p_2^{-1} )
 \\ q_2, \, p_2^{-1}
\end{array};
\, \ldots \, ;\right. }
\\ \\
\phantom{aaaaaaaaaaaaaaaaaaaaaaaaaaaa}\displaystyle{\left.
\begin{array}{c}\beta_{N}(q_N^{-1}\! -\! q_N )
\\ p_{N}, \, q_{N}^{-1}\end{array};
\begin{array}{c}\beta_{N} (p_N\! -\! p_N^{-1} )\\ 
q_{N}, \, p_{N}^{-1}\end{array}
\right ]
({\bf T}, -{\bf T})},
\end{array}
\eeq
where $\beta_i$, $p_i$, $q_i$ are $3N$ arbitrary parameters.
For example, at $N=1$ with $p_1=p$, $q_1=q$, $\beta_1=\beta$ 
we have
\beq\label{s4a}
\begin{array}{c}
\displaystyle{
\tilde \tau_n({\bf T}, -{\bf T})=
1+\beta (p\! -\! p^{-1}\! +\! q^{-1}\! -\! q)(pq)^n e^{\zeta ({\bf T})}
+\beta^2 (p-q)^2(pq)^{2n-1}
e^{2\zeta ({\bf T})}}
\\ \\
=\Bigl (1+\beta (p-q)(pq)^n e^{\zeta ({\bf T})}\Bigr )
\Bigl (1+\beta (p-q)(pq)^{n-1} e^{\zeta ({\bf T})}\Bigr ),
\end{array}
\eeq
where
\beq\label{s5}
\zeta ({\bf T})=\xi ({\bf T}, p)+\xi ({\bf T}, q)-
\xi ({\bf T}, p^{-1})-\xi ({\bf T}, q^{-1}).
\eeq
Therefore, the one-soliton 
tau-function $\tau_n^B({\bf T})$ has the form
\beq\label{s6}
\tau_n^B({\bf T})=\exp \Bigl (\frac{1}{2}\sum_{k\geq 1}kT_k^2\Bigr )
\Bigl (1+\beta (p-q)(pq)^n e^{\zeta ({\bf T})}\Bigr ).
\eeq

Similar but longer calculations lead to the following result for the 
2-soliton tau-function:
\beq\label{s7}
\begin{array}{l}
\displaystyle{
\tau_n^B({\bf T})=
\exp \Bigl (\frac{1}{2}\sum_{k\geq 1}kT_k^2\Bigr )
\Bigl (1+\beta_1 (p_1\! -\! q_1)(p_1q_1)^n e^{\zeta _1({\bf T})}+
\beta_2 (p_2\! -\! q_2)(p_2q_2)^n e^{\zeta _2({\bf T})}}
\\ \\
\phantom{aaaaaaaaaaaaaaaaaaaaaaaaaaaa}
+\beta_1\beta_2 (p_1\! -\! q_1)(p_2\! -\! q_2)b_{12}
e^{\zeta _1({\bf T}) +\zeta _2({\bf T})}\Bigr ),
\end{array}
\eeq
where
$$
\zeta_i ({\bf T})=\xi ({\bf T}, p_i)+\xi ({\bf T}, q_i)-
\xi ({\bf T}, p_i^{-1})-\xi ({\bf T}, q_i^{-1})
$$
and
\beq\label{s8}
b_{ij}=\frac{(p_i-q_j)(p_i-p_j)(q_i-p_j)(q_i-q_j)}{(p_iq_j\! -\! 1)
(p_ip_j\! -\! 1)(q_ip_j\! -\! 1)(q_iq_j\! -\! 1)}.
\eeq

Let us show that the $N$-soliton tau-function has the general form
\beq\label{s9}
\tau_n^B({\bf T})=
\exp \Bigl (\frac{1}{2}\sum_{k\geq 1}kT_k^2\Bigr )
\sum_{m=0}^{N}\,
\sum_{i_1<i_2< \ldots < i_m}^{N} \Bigl (\prod_{\gamma =1}^m 
\beta_{i_{\gamma}}(p_{i_{\gamma}}\! -\!  q_{i_{\gamma}})(p_{i_{\gamma}}
q_{i_{\gamma}})^n
e^{\zeta_{i_{\gamma}}({\bf T})}\Bigr )
\prod_{\mu <\nu}^{m}b_{i_{\mu}i_{\nu}}.
\eeq
To see this, we will show that the functions (\ref{s4}) and
(\ref{s9}) satisfy equation (\ref{tau2}). 
As a direct calculation shows, this fact
is based on an identity for rational functions. Let 
$I$ be the set $\{1,2, \ldots , N\}$ and $I_1, I_2$ subsets of it
such that $I_1 \cup I_2 =I$, $I_1\cap I_2 =\emptyset$. The identity
is
\beq\label{s10}
\begin{array}{l}
\displaystyle{
\prod_{i\in I} (p_i-q_i)\sum_{I_1, I_2}\prod_{i\in I_1}
\frac{(z- p_i)(z- q_i)}{(p_iz\! -\! 1)(q_iz\! -\! 1)}
\!\! \prod_{{\scriptsize 
\begin{array}{c} i,j\in I_1\\ i<j\end{array}}}\!\! 
\frac{(p_i-p_j)(p_i-q_j)(q_i-q_j)(q_i-p_j)}{(p_ip_j\! -\! 1)(p_iq_j\! -\! 1)
(q_iq_j\! -\! 1)(q_ip_j\! -\! 1)}
}
\\ \\
\phantom{aaaaaaaaaaa}\displaystyle{\times  
\prod_{{\scriptsize 
\begin{array}{c} i,j\in I_2\\ i<j\end{array}}}\!\! 
\frac{(p_i-p_j)(p_i-q_j)(q_i-q_j)(q_i-p_j)}{(p_ip_j\! -\! 1)(p_iq_j\! -\! 1)
(q_iq_j\! -\! 1)(q_ip_j\! -\! 1)}
}
\\ \\
\phantom{aaaaaaaaaaa}
\displaystyle{=\, \sum_{I_1, I_2}\Bigl (\prod_{i\in I_1}
\frac{z-p_i}{q_iz-1}\, (1-q_i^2)\Bigr )
\Bigl (\prod_{i\in I_2}
\frac{z-q_i}{p_iz-1}\, (p_i^2-1)\Bigr )
}
\\ \\
\displaystyle{\times \prod_{{\scriptsize 
\begin{array}{c} i,j\in I_1\\ i<j\end{array}}}\!\! 
\frac{(p_i-p_j)(q_i-q_j)}{(p_iq_j\! -\! 1)
(q_ip_j\! -\! 1)}
\!\! \prod_{{\scriptsize 
\begin{array}{c} i,j\in I_2\\ i<j\end{array}}}\!\! 
\frac{(p_i-p_j)(q_i-q_j)}{(p_iq_j\! -\! 1)
(q_ip_j\! -\! 1)}
\prod_{i_1\! \in \! I_1, \, i_2\! \in \! I_2} \!
\frac{(p_{i_1}-p_{i_2})(q_{i_1}-p_{i_2})}{(p_{i_1}p_{i_2}\! -\! 1)
(q_{i_1}q_{i_2}\! -\! 1)}.}
\end{array}
\eeq
The sums are taken over all unions of the set $I$ into two disjoint
subsets $I_1, I_2$ ($I_1$ or $I_2$ may be empty). The left (right )
hand
side of the identity 
comes from the right (left) hand side of (\ref{tau2}).
It is not difficult
to see that this identity is true for $N=1$. The proof in general case
is by induction. Suppose (\ref{s10}) holds for some $N$, then the following
argument shows that it holds for $N+1$. The both sides of (\ref{s10})
are rational functions of $z$ with $2N$ simple poles at $z=p_i^{-1}$ and 
$z=q_i^{-1}$ such that they are $O(1)$ as $z\to \infty$. 
Consider, for example, the poles at $z=p_1^{-1}$. It is not difficult
to see that equality of the residues at this pole is equivalent to
the same identity (\ref{s10}) but with $N\to N-1$ and $z=q_1$. Therefore,
the residues of both sides are equal due to the induction assumption.
The other poles can be considered in a similar way. This allows us 
to conclude that the difference of the both sides of (\ref{s10}) does
not depend on $z$. In order to be sure that it is equal to zero, it is
enough to calculate it for some particular value of $z$, say $z=p_1$. 
Again, the induction assumption implies that it is indeed equal to zero.

Finally, let us note that the tau-function (\ref{s9}) is a pfaffian.
Recall that pfaffian of a $2N\! \times \! 2N$ antisymmetric matrix
$A$ is square root of its determinant (which is a full square). It 
is given by the explicit formula
$$
\mbox{pf}\, A =\sum (-1)^P A_{i_1i_2}A_{i_3i_4}\ldots A_{i_{2N-1}i_{2N}},
$$
where the sum is taken over all permutations of indices 
$1,2, \ldots , 2N$ such that $$i_1<i_2, \, 
i_3<i_4, \ldots , i_{2N-1}<i_{2N}, \qquad 
i_1<i_3<\ldots < i_{2N-1}$$ 
and $P$ is parity of the permutation. For example, at $N=1$ we have
$\mbox{pf}\, A =A_{12}$ and at $N=2$
$\mbox{pf}\, A =A_{12}A_{34}-A_{13}A_{24}+A_{14}A_{23}$. 

Let $J$ be the  $2N\! \times \! 2N$ antisymmetric matrix with matrix
elements $J_{2i-1, 2i}=1$, $J_{2i, 2i-1}=-1$ and $0$ otherwise. 
Let $B$ be the following antisymmetrix matrix:
\beq\label{s13}
B_{ij}=J_{ij}+\nu_i \nu_j \frac{r_i-r_j}{r_ir_j-1}\, (r_ir_j)^n
e^{\zeta ({\bf T}, r_i)+\zeta ({\bf T}, r_j)},
\eeq
where
$r_{2i-1}=p_i$, $r_{2i}=q_i$, 
$\nu_{2i-1}=\nu_{2i}=\gamma_i$ ($i=1, \ldots , N$),
$$
\zeta ({\bf T}, r)=\xi ({\bf T}, r)-\xi ({\bf T}, r^{-1}).
$$
Then the tau-function (\ref{s9}) with
$\displaystyle{\beta_i=\frac{\gamma_i^2}{p_iq_i-1}}$ is equal to
\beq\label{s14}
\tau_n^B({\bf T})=
\exp \Bigl (\frac{1}{2}\sum_{k\geq 1}kT_k^2\Bigr )
\, \mbox{pf}\, (J+B).
\eeq
The equivalence of (\ref{s9}) and (\ref{s14}) 
is based on the following formula 
\beq\label{s11}
\mbox{pf}\left (\frac{x_i-x_j}{x_{i}x_{j}-1}\right )_{1\leq i,j\leq 2n}=
\prod_{k<l}\frac{x_k-x_l}{x_kx_l-1}
\eeq
(see, for example, \cite{IW95,IW99}) which can be regarded as
a pfaffian analog of the 
formula for determinant of the Cauchy matrix and on the formula for
pfaffian of the matrix $J+B$:
\beq\label{s12}
\mbox{pf}\, (J+B)=1+ \sum_{m=1}^{N}\sum_{i_1< \ldots <i_m}^N
\mbox{pf}\, B(2i_1\! -\! 1, 2i_1; 2i_2\! -\! 1, 
2i_2;\ldots ; 2i_m\! -\! 1, 2i_m),
\eeq 
where $B(2i_1\! -\! 1, 2i_1; 2i_2\! -\! 1, 
2i_2;\ldots ; 2i_m\! -\! 1, 2i_m)$ is the antisymmetric $2m\! \times \! 2m$
matrix consisting of rows and columns of matrix $B$ numbered by 
$2i_1\! -\! 1, 2i_1; 2i_2\! -\! 1, 
2i_2;\ldots ; 2i_m\! -\! 1, 2i_m$. This latter formula can be proved 
by induction in $N$. 

\section{Concluding remarks}

In this paper we have reconsidered the B-Toda hierarchy which is 
the subhierarchy of the Toda lattice obtained from it by imposing
the constraint (\ref{int1}) or (\ref{b2}) (the constraint of type B). 
Our main concern in this paper was formulation of the B-Toda hierarchy
in terms of the tau-function, which remained obscure in the first work
\cite{KZ22} on the subject. We have characterized tau-functions of the
Toda lattice $\tau_n({\bf t}, \bar {\bf t})$ 
such that they provide solutions to the B-Toda hierarchy
(the condition (\ref{b9})). Besides, we have introduced the 
tau-function $\tau^B_n({\bf T})$ of the B-Toda hierarchy depending 
on the independent variables $T_k=\frac{1}{2}(t_k-\bar t_k)$ (with
$t_k+\bar t_k=0$). The existence of this tau-function follows from the
bilinear equation (\ref{tau9}) for the wave functions (solutions to the
auxiliary linear problems). We have also obtained the integral 
bilinear relation for the tau-function $\tau^B_n({\bf T})$ which 
encodes all equations of the hierarchy. An important consequence 
of this relation is the 4-term bilinear equation (\ref{tau11}) for the 
$\tau^B_n({\bf T})$ which after a linear change 
of variables is shown to be equivalent to the fully discrete version
of the BKP equation written in terms of the tau-function $\tau^{BKP}$.

We also considered in detail tau-functions of soliton solutions 
to the B-Toda lattice hierarchy and showed that they are expressed as
pfaffians of certain antisymmetric matrices. This is similar to what
was known for soliton solutions of the BKP hierarchy \cite{Hirota,HT96}.

\section*{Acknowledgments}

\addcontentsline{toc}{section}{Acknowledgments}

This work 
was supported by the 
Russian Science Foundation under grant 19-11-00062, project
https://rscf.ru/en/project/19-11-00062/.

\end{document}